# MonteCarloMeasurements.jl: Nonlinear Propagation of Arbitrary Multivariate Distributions by means of Method Overloading

Fredrik Bagge Carlson *

January 21, 2020

## Abstract

This manuscript outlines a software package that facilitates working with probability distributions by means of Monte-Carlo methods, in a way that allows for propagation of multivariate probability distributions through arbitrary functions. We provide a *type* that represents probability distributions by an internal vector of unweighted samples, `Particles`, which is a subtype of a `Real` number and behaves just like a regular real number in calculations by means of method overloading. This makes the software easy to work with and presents minimal friction for the user. We highlight how this design facilitates optimal usage of SIMD instructions and showcase the package for uncertainty propagation through an off-the-shelf ODE solver as well as for robust probabilistic optimization with automatic differentiation.

## 1 Introduction

Technical computing involving quantities distributed according to some probability distribution is important in most fields of science and engineering. A probability distribution can represent both the probability of of some future event or outcome, or the uncertainty associated with a quantity such as a measurement or inferred parameter. A variable or parameter might be associated with uncertainty if it is, e.g., measured and subject to measurement error, or otherwise estimated from data. While performing computations with deterministic values is straightforward, doing the same with values with an associated probability distribution is, outside a few special cases, highly nontrivial. One such special case is when an affine function $y = f(x) = Ax + b$ is applied to a normally distributed variable $x \sim N(\mu, \Sigma)$. The posterior distribution associated with the random variable $y$ will remain normal: $y \sim N(A\mu + b, A\Sigma A^\mathsf{T})$. This fortunate fact underpins the effectiveness of methods like celebrated Kalman filter for dynamic filtering and the least-squares method for parameter estimation. When the function applied to a random variable is nonlinear, or the random variable is not normally distributed, the posterior distribution usually lacks a closed-form expression, making it significantly harder to work with, reason about and performing computations with.

Proper calculation of the distribution of $y$ requires numerical computation of integrals. While this is feasible in low dimensions, it quickly becomes intractable as the dimension increases, necessitating approximative approaches. Two common approximation techniques have emerged. The first approach is to linearize the function and approximate the prior distribution with a normal distribution, after which one can perform linear uncertainty propagation of normal distributions. The second approach is to approximate the prior distribution with samples, each of which can be propagated through the function individually to form a collection of samples to represent the posterior distribution. This approach is often referred to as the Monte-Carlo method and is illustrated in Fig. 1.

These two methods have different drawbacks. While the Monte-Carlo method is rather simple to implement, the linearization approach requires some form of support by automatic differentiation to alleviate the the requirement for the user to provide the Jacobian of the function to propagate uncertainty through. The Monte-Carlo method can approximate arbitrary prior and posterior distributions and handle highly nonlinear functions, but may require many samples to yield accurate results in high dimensions, increasing the computational cost. The linearization approach may be more computationally efficient, but may result in false inferences for highly nonlinear functions. Software tools assisting in applying these two methods to uncertainty propagation are plentiful, a list of which is available at Wikipedia: "List of uncertainty propagation software".

In this manuscript, we describe the implementation of MonteCarloMeasurements.jl, a software package written in the Julia programming language (Bezanson et al., 2017) that uses the multiple-dispatch paradigm of Julia to provide numerical types that behave like regular numbers, but internally represent and propagate sample-based representations of probability distributions. The powerful method dispatch system of Julia allows for uncertainty-propagation that both presents minimal friction for the user, while also allowing optimal usage of the SIMD-processing[1] units of modern processors, significantly mitigating the computational cost of Monte-Carlo evaluation of a program.

## 2 MonteCarloMeasurements.jl

As alluded to above, MonteCarloMeasurements.jl facilitates working with probability distributions by providing special numerical uncertainty types for which methods of standard functions have been implemented. To make this statement more concrete, consider a program that operates on an input $x$ and is composed of calls to functions (ond operators) like `+,-,*,/,sin,exp` etc. This program may have been written by the user or come from a third party library, as long as it is written in Julia. If this program accepts and operates on any form of real number, it will also operate on the provided uncertainty type since methods of `+,-,*,/,sin,exp` have been implemented for this type and Julia will automatically dispatch to these methods when the program is executed with an input $x$ of the uncertainty type.

The main type provided by the package is called `Particles{T,N}`, which is parameterized by the number of samples $N$ and the data type of the samples $T$. An instance of the type `Particles`[2] represents the probability distribution of a variable by a vector of unweighted samples and is a subtype of the abstract number type `Real`. `Particles` can thus be used and propagated through any Julia program that accepts standard real numbers,[3] such as integers or floating point numbers.

---

*The author would like to acknowledge Chad Scherrer, the author of Soss.jl, Simeon Schaub, Seth Axen and Mason Protter for contributions to MonteCarloMeasurements.jl and the surrounding ecosystem as well as Mosè Giordano, the author of Measurements.jl, which served as an inspiration for MonteCarloMeasurements.jl. baggepinnen@gmail.com. Open-source implementation available at github.com/baggepinnen/MonteCarloMeasurements.jl

[1] Single Instruction Multiple Data.
[2] The name "particles" comes from the particle-filtering literature, where a particle is synonymous with a sample.

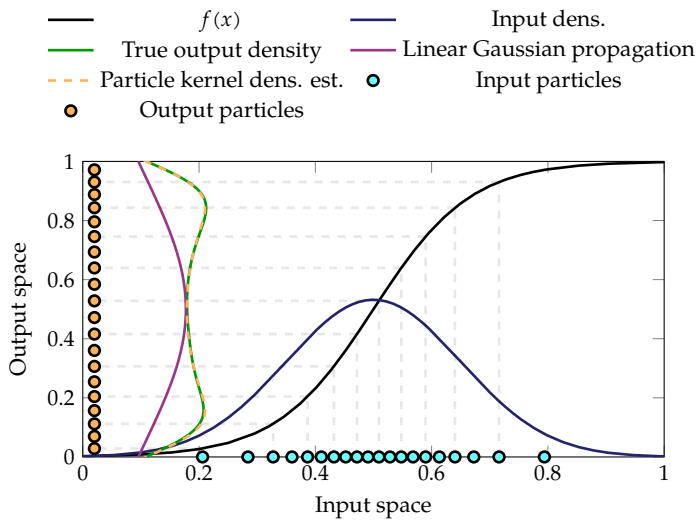

Figure 1: Uncertainty propagation through a function $f(x)$: Particles, depicted as cyan points along the horizontal axis, represent the probability distribution of the input (blue). The function $f$ (black) is evaluated at each input point to produce a set of output points, shown along the vertical axis (yellow). These points form an approximation to the true probability distribution of the output (green). Using kernel-density estimation, a continuous approximation to the particle cloud is shown (yellow dash). Linear uncertainty propagation produces the purple output distribution.

A variable $p$ of type `Particles` behaves just as any other number while partaking in calculations. Internally, every function applied to $p$ is dispatched to a method that applies the function to every sample in the vector of samples stored in $p$. After a calculation, an approximation to the complete posterior distribution of the output is captured and represented by the output particles.

## 2.1 Basic Usage

We illustrate the look-and-feel of the interface for creating uncertain parameters below

```julia
julia> using MonteCarloMeasurements, Plots, Distributions
julia> a = π ± 0.1 # Construct Gaussian uncertain parameters using ± (\pm)
Part500(3.142 ± 0.1)
julia> b = 2 ∓ 0.1 # ∓ (\mp) creates StaticParticles (with StaticArrays)
SPart100(2.0 ± 0.1)
julia> std(a)      # Ask about statistical properties
0.09997062445203879
julia> sin(a)      # Use them like any real number
Part500(1.255e-16 ± 0.0995)
julia> sin(a)/cos(a) - tan(a) # Self-correlation is naturally handled
Part500(0.0)
julia> plot(a)     # Plot them
julia> b = sin.(1:0.1:5) .± 0.1; # Create multivariate uncertain numbers
julia> plot(b)     # Vectors of particles can be plotted
julia> c = Particles(500, Poisson(3.)) # Create uncertain numbers distributed
↪      according to a given distribution
Part500(2.896 ± 1.71)
```

---

[3] Complex numbers in julia are simply structures with two real numbers, allowing particles to work with complex numbers through the construct `Complex{Particles}`.

## 2.2 Interaction with the Julia Ecosystem

The mean, standard deviation, quantile etc. can be extracted from particles using the corresponding functions. Particles also interact with Distributions.jl (Lin et al., 2019), so that a call like, e.g., `Normal(p)` will return a `Normal` type from Distributions.jl. Plot recipes for Plots.jl are provided so that particles and vectors of particles can be easily visualized and ArviZ.jl provides further visualization support for particles. Functions with internal uncertain parameters and with uncertain inputs can be differentiated using the automatic-differentiation library Zygote.jl (Innes, 2018).

## 2.3 Performance Advantages

Monte-Carlo evaluation of a computer program lends itself very well to parallel execution, since each invokation of the program is independent. While parallel execution is supported by MonteCarloMeasurements.jl, we will highlight two aspects of the software that is far more difficult to achieve, and that yield increased performance even in the absence of parallel computing cores.

### 2.3.1 Shared computations

When the function to propagate uncertainty through spends significant time performing computations that are unaffected by the uncertain variables, the naive approach to Monte-Carlo evaluation suffers from performing this computation for each invokation of the function. Consider, e.g., the following program

```julia
function least_squares(A,y)
    Q = qr(A)
    return Q\y
end
```

if only the input variable $y$ is uncertain, the QR-factorization of $A$ can be done once only, while naive application of the Monte-Carlo method would perform it $N$ times for $N$ samples. Using MonteCarloMeasurements.jl, this function would be invoked once only, with a matrix $A$ of regular floating-point values and a vector $y$ of particles. Thus, only the backsolve would be carried out $N$ times.

Further, any dynamic dispatch occurring in the program will, using the MonteCarloMeasurements approach, be paid for once only, whereas repeated invokation of the function will pay the price for dynamic dispatch once per invokation.

### 2.3.2 Optimal usage of SIMD instructions

Modern processors contain processing units that are able to execute the same instruction on multiple input data simultaneously (on a single CPU core). These instructions, referred to as Single Instruction Multiple Data (SIMD), can typically operate on 4-32 values at the same time, depending on the bit width of the values and the processor architecture. Unfortunately, many programs do not lend themselves to application of the SIMD instructions, a notable example being loops where each iteration depends on previous iterations. However, multiple such loops can executed in parallel and very well utilize the SIMD instructions to increase the amount of data processed by each instruction. Consider, for instance, the simulation of an ordinary differential equation (ODE), where some of the parameters or the initial condition are uncertain. Naive application of the Monte-Carlo method would not be able to utilize the SIMD instructions of the processor since loop iterations are dependent. While propagating uncertainties using MonteCarloMeasurements.jl, each primitive operation carried out by the ODE solver is performed over the entire vector of samples simultaneously, making it trivial for the compiler to utilize SIMD instructions for close to every instruction in the entire simulation.

An example of ODE simulation is provided in Sec. 3.1.

## 2.4 Systematic Sampling

The variance introduced by Monte-Carlo sampling has some fortunate and some unfortunate properties. It decreases as $1/N$, where $N$ is the number of particles/samples. This unfortunately means that to reduce the standard deviation in an estimate, one must quadruple the number of particles. On the other hand, this variance does not depend on the dimension of the space, which is fortunate and allows application of the Monte-Carlo method to high-dimensional problems.

In MonteCarloMeasurements.jl, we perform systematic sampling (Douc, Cappé, and Moulines, 2005) whenever possible. This approach exhibits lower variance than standard random sampling. Systematic sampling makes use of the quantile function (inverse cdf) to distribute samples in a way that minimizes the discrepancy while exactly obeying the target distribution.

## 2.5 Multivariate Distributions

MonteCarloMeasurements.jl natively handles multivariate distributions—a multivariate random variable is simply represented as an array of Particles. The constructor Particles(N, d::Distribution) will, if $d$ is a multivariate distribution, return an array of particles where particles exhibit the desired correlation.

Below, we illustrate the creation of a bivariate random variable $p$, which we transform using the linear transform $y = Ap$. We then compare the covariance of the resulting array of particles $y$ to the theoretical covariance matrix $A\Sigma A^\mathsf{T}$.

```
julia> p = [1 ± 1, 5 ± 2] # Create a vector of uncorrelated particles
2-element Array{Particles{Float64,500},1}:
 1.0 ± 1.0
 5.0 ± 2.0

julia> A = randn(2,2); # Create a random matrix

julia> y = A*p # Transform particle vector
2-element Array{Particles{Float64,500},1}:
 -8.04 ± 3.1
  2.4 ± 1.5

julia> cov(y) # Covariance of posterior multivariate particles
2×2 Array{Float64,2}:
  9.61166  -3.59812
 -3.59812   2.16701

julia> A*Diagonal([1^2, 2^2])*A' # Theoretical posterior covariance
2×2 Array{Float64,2}:
  9.4791   -3.53535
 -3.53535   2.15126
```

### 2.5.1 Sigma Points and the Unscented Transform

An intermediate step between pure Gaussian propagation and the Monte-Carlo method is something referred to as the unscented transform (Menegaz et al., 2015). The transform amounts to selecting a small set of points, referred to as *sigma points*, that have a pre-specified mean and variance. These samples are then propagated through a function and can, in the case of the unscented transform, be used to fit a normal distribution to the output points. MonteCarloMeasurements.jl supports creating sigma points for uncertainty propagation as they provide a compromise between fidelity and computational cost. The relative performance between various uncertainty-propagation methods is showcased in Sec. 3.1.

| Uncertainty propagation | Dynamic filtering | Method |
|---|---|---|
| Linear/Gaussian | Extended Kalman filter | Linearization |
| Particles(sigmapoints) | Unscented Kalman filter | Unscented transform |
| Particles | Particle filter | Monte Carlo |

Table 1: Relation between uncertainty propagation and dynamic filtering.

Uncertainty propagation is intimately linked with dynamic filtering methods such as the particles filter and unscented Kalman filter, the relation between methods for uncertainty propagation and dynamics filtering is emphasized in Table 1.

## 2.6 Limitations

While the multiple-dispatch paradigm of Julia is very powerful, there is one particular construct that is hard to handle: control flow where the branch is decided by an uncertain value. Consider the following case

```
function negsquare(x)
    x > 0 ? x^2 : -x^2
end
p = 0 ± 1
```

Ideally, half of the particles should turn out negative and half positive when applying negsquare(p). However, this will not happen as the $x > 0$ is not defined for uncertain values. In this simple case, we can easily circumvent this problem by registering the function negsquare as a primitive, after which particles will be propagated one by one through the entire function. However, if the branching condition involves particles stored as the field of a struct, it becomes much harder to circumvent the problem using dispatch. While this problem is fundamental to the dispatch-based method of uncertainty propagation, it is not fundamental to the Julia programming language. A proof-of-concept software package implementing a compiler transform that would allow for propagation of particles through arbitrary control flow exists, but is as of the time of writing not yet mature.[4]

A possible workaround for the above mentioned limitation is to include type parameters in the structure containing the Particles such that this case can be dispatched for, and providing a special method of the function containing the uncertain control flow.

# 3 Usage Examples and Benchmarks

## 3.1 Uncertainty Propagation—ODE simulation

This example is based on a tutorial[5] for simulating an ODE using OrdinaryDiffEq.jl (Rackauckas and Nie, 2017) with the linear uncertainty-propagation package Measurements.jl (Giordano, 2016). Consider a pendulum with dynamics described by

$$\dot{u}_1 = \dot{\theta}$$
$$\dot{u}_2 = -\frac{g}{L}\sin(\theta)$$
$$g = 9.79 \pm 0.02$$
$$L = 1.00 \pm 0.01$$
$$u_0 = [0 \pm 0, \pi/3 \pm 0.02]$$

where $\mu \pm \sigma \sim N(\mu, \sigma^2)$ denotes a normally distributed quantity. The dynamics of the pendulum have uncertainties in both the gravitational-acceleration constant $g$, the length of the pendulum $L$ and the initial condition $u_0$. If this system is simulated for a short duration, linear and Monte-Carlo uncertainty propagation produces very similar uncertainty estimates, shown in the left panel of Fig. 2. If integration is carried out over a longer horizon, the errors incurred by linearization compounds and result in very unreliable results, shown in the right panel of Fig. 2. The linear approach leads to the conclusion that the amplitude at some points might have increased to much higher than the starting amplitude, implying that energy somehow has been added to the system. Using the Monte-Carlo method, each trajectory has a constant amplitude (although individual trajectories amplitudes vary slightly due to the uncertainty in the initial angle), but the phase is completely uncertain due to the slightly different frequencies resulting from the uncertainty in the pendulum length.

---
[4] github.com/FluxML/Hydra.jl
[5] tutorials.juliadiffeq.org/html/type_handling/02-uncertainties.html

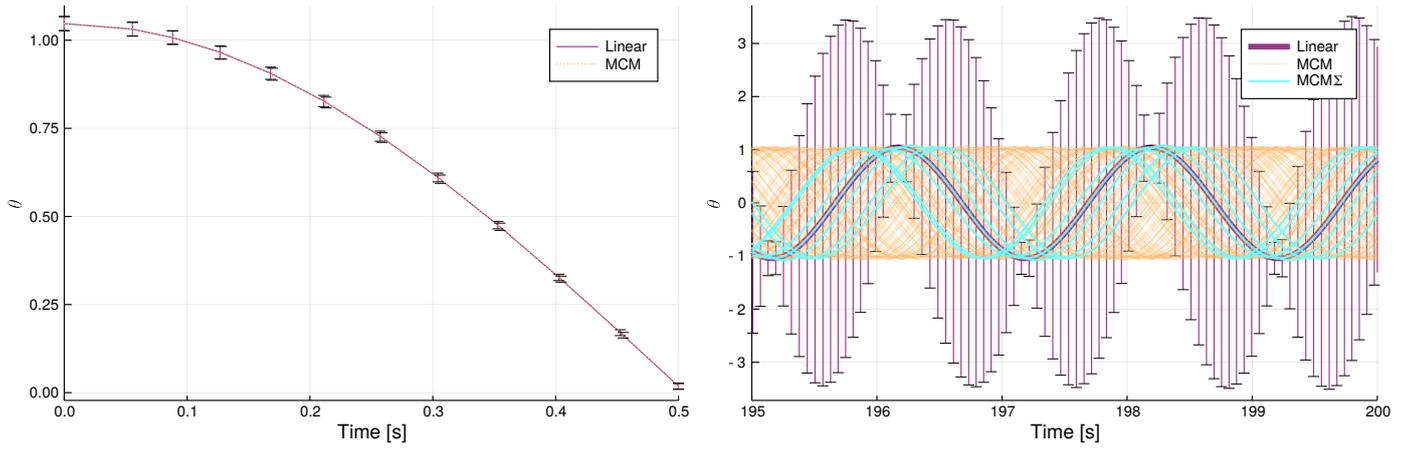

Figure 2: ODE simulation. The left panel shows the angle of the pendulum while simulating for 0.5 seconds using both linear uncertainty propagation, performed using Measurements.jl, and Monte-Carlo propagation using MonteCarloMeasurements.jl. Both methods produce similar uncertainty estimates. The right panel shows the result for after simulating for 195 seconds. When simulating for this long, the errors incurred by the linearization have grown so large that the result is unreliable. The black lines show 100 realizations using Monte-Carlo uncertainty propagation and the green lines show simulation using 7 sigma points.

|  | Float32 | Linear | MCM | MCM Σ | Naive MC |
|---|---|---|---|---|---|
| Time [ms] | 0.7 | 9.9 | 12.9 | 2.7 | 88.7 |
| Memory [MB] | 1.2 | 30.0 | 30.7 | 5.6 | 119.6 |
| k Allocations | 12.2 | 691.2 | 59.6 | 61.2 | 1221.9 |

Table 2: Computational cost of simulating the ODE benchmark problem for 100 seconds. Linear uncertain propagation is performed using the library Measurements.jl. Monte-Carlo evaluation was done using 100 samples using both MonteCarloMeasurements.jl (MCM) and the naive method. MCM Σ denotes MC evaluation using 7 sigma points.

The computational costs associated with the various simulations are shown in Table 2. Simulation using only 32 bit floating-point values is about 14 times faster than linear uncertainty propagation and 19 times faster than MonteCarloMeasurements.jl using 100 samples. Interestingly, naively simulating 100 times using Float32 is almost 7 times slower than MonteCarloMeasurements.jl using the same number of samples, illustrating how MonteCarloMeasurements.jl facilitates the usage of SIMD instructions. If the simulation is performed using 7 sigma points ($2n + 1$ where $n = 3$ is the number of uncertain parameters), the result is obtained in only 4 times the execution time of a single simulation with Float32, making nonlinear uncertainty propagation very computationally affordable and 4 times faster than linear uncertainty propagation.

To illustrate usage of MonteCarloMeasurements.jl, we reproduce the code required to simulate the pendulum in Algorithm 1. The only modification to this code required to add uncertainty propagation is the

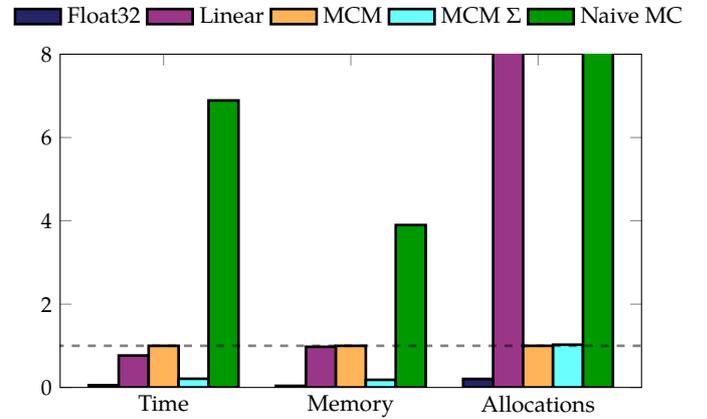

Figure 3: Depiction of the performance result presented in Table 2. Results are normalized so that the value for MCM.jl with 100 particles is 1.

**Algorithm 1** Code to simulate the pendulum with uncertain parameters. The operator ± creates Particles{Float64,500} with the specified mean and standard deviation.

```
using MonteCarloMeasurements, OrdinaryDiffEq, Plots

g  = 9.79 ± 0.02 # Gravitational constant
L  = 1.00 ± 0.01 # Length of the pendulum
u₀ = [0.0 ± 0.0, π / 3.0 ± 0.02] # Initial speed and initial angle

function pendulum(u̇,u,p,t)
    θ, θ̇ = u[1], u[2]
    u̇[1] = θ̇
    u̇[2] = -(g/L)sin(θ)
end

prob = ODEProblem(pendulum, u₀, (0.0, 2.0))
sol  = solve(prob, Tsit5(), reltol = 1e-6)
plot(sol)
```

addition of the ± operator that creates Particles instead of floating-point parameters, the ODE solver library and the plotting library are oblivious to MonteCarloMeasurements.jl.

## 3.2 MCMC posterior

Bayesian inference using Markov Chain Monte Carlo (MCMC) methods naturally produces a sample-based representation of the posterior density. The probabilistic programming package Soss.jl[6] employs MonteCarloMeasurements.jl for convenient handling of the inference result in the form of Particles. Using Particles, one can easily form predictions and "push distributions through the model". Additionally, the package ArviZ.jl[7] implements support for sophisticated visualization of Bayesian models and has native support for Particles.

## 3.3 Robust Optimization

In Algorithm 2, we showcase how MonteCarloMeasurements.jl can be used to solve a robust optimization problem. We specify a cost function containing uncertain parameters and use Zygote.jl (Innes, 2018) to obtain a gradient function (cost' returns a function that returns the gradient of cost) and Optim.jl (Mogensen and Riseth, 2018) to minimize the cost function. Using Zygote.jl, it is possible to differentiate functions containing uncertain parameters, with respect to uncertain inputs, giving the user great flexibility in specifying and solving robust and probabilistic optimization problems.

The documentation of MonteCarloMeasurements.jl contains several more detailed examples of robust optimization, including robust PID-controller optimization.

**Algorithm 2** Robust-optimization example. In the cost function below, we ensure that $cx + dy > 10 \ \forall \ c, d \in P$ by looking at the worst case.

```julia
using MonteCarloMeasurements, Optim, Zygote
const c = 1 ∓ 0.1 # These are the uncertain parameters
const d = 1 ∓ 0.1 # These are the uncertain parameters

function cost(pars)
    x,y = pars
    -(3x+2y) + 10000*(maximum(c*x+d*y) > 10)
end

pars = [1., 1] # Initial guess
res = Optim.optimize(cost, cost', pars, BFGS(), inplace=false)
```

## 4 Concluding Remarks

The multiple-dispatch paradigm makes it a joy to implement types that interact with the whole Julia ecosystem. This package even provides a type of particles that execute all operations on the GPU with the help of the package CuArrays.jl (Besard et al., 2019). The GPU support is currently in beta and preliminary benchmarks currently show a modest factor of 5-7 times performance benefit in the ODE benchmark for sample sizes of 100 000–1 000 000. Future improvements to the memory management in the Julia GPU ecosystem is likely to improve upon this number.

The infrastructure provided by MonteCarloMeasurements.jl can also be used to accelerate population-based optimization algorithms such as Particle Swarm Optimization and particle filters, by allowing GPU or SIMD-optimized execution of arbitrary functions that are oblivious to such techniques.

---

[6] github.com/cscherrer/Soss.jl
[7] github.com/arviz-devs/ArviZ.jl

Code to reproduce the figures and examples in this paper is provided in the software repository github.com/baggepinnen/MonteCarloMeasurements.jl. MonteCarloMeasurements.jl and all code in this paper is licensed under the MIT license.